\documentclass[aps]{revtex4}

\usepackage{amssymb}
\usepackage{amsmath}
\usepackage{graphicx}
\usepackage{bmpsize}
\usepackage[a4paper,  margin=1in]{geometry}
\usepackage{longtable}

\begin{document}

\title{Phenomenology of coupled non linear oscillators}

\author{E. \surname{Estevez-Rams}}
\email{estevez@fisica.uh.cu}
\affiliation{Facultad de F\'isica-Instituto de Ciencias y Tecnolog\'ia de Materiales(IMRE), Universidad de La Habana, San Lazaro y L. CP 10400. La Habana. Cuba.}

\author{D. \surname{Estevez-Moya}}
\affiliation{Facultad de F\'isica, Universidad de La Habana, San Lazaro y L. CP 10400. La Habana. Cuba.}

\author{B. \surname{Arag\'on-Fern\'andez}}
\affiliation{Universidad de las Ciencias Inform\'aticas (UCI), Carretera a San Antonio, Boyeros. La Habana. Cuba.}

\date{\today}

\begin{abstract}
A recently introduced model of coupled non linear oscillators in a ring is revisited in terms of its information processing capabilities. The use of Lempel-Ziv based entropic measures allows to study thoroughly the complex patterns appearing in the system for different values of the control parameters. Such behaviors, resembling cellular automata, have been characterized both spatially and temporally. Information distance is used to study the stability of the system to perturbations in the initial conditions and in the control parameters. The latter is not an issue in cellular automata theory, where the rules form a numerable set, contrary to the continuous nature of the parameter space in the system studied in this contribution. The variation in the density of the digits, as a function of time is also studied.  Local transitions in the control parameter space are also discussed.
\end{abstract}

\maketitle

\section{Introduction}

Coupled non linear oscillators have been studied at least since the sixties of the last century, as they model a number of interesting physical, biological, and chemical systems \cite{winfre67,kuramoto75,strogatz01,mosekilde02,zillmer06}. Winfree \cite{winfre67} introduced the notion that under weak coupling conditions, amplitude variations in the oscillators can be neglected and instead, the study of the phases display the relevant dynamic of the system. Phase equations are then derived as appropriate representations of the oscillatory dynamics. Within this approximation the Kuramoto model is probably the most studied \cite{kuramoto75}. In the Kuramoto model, non linear, almost identical, oscillators are weakly coupled with a sinusoidal interaction that depends on the phase difference between pair of oscillators. The coupling is global as each oscillator can interact with all the others in the system. Kuramoto model has been used as a convenient model to study synchronization between coupled units ( See \cite{acebron05} and references therein).The common approach is to assign a  natural frequency to each oscillator which can be affected by an ''intrinsic'' or self feedback coming from the same oscillator, and an interaction term that comes from the coupling to the other oscillators.

Understanding the emergence of new phenomena from the collective interaction of parts of a system is one of the main quest of non linear dynamics and related areas \cite{crutchfield12}. In the case of an array of coupled oscillators the question has been posed relevant to theoretical biology and the workings of neural networks at the brain cortex \cite{ermentrout98,mosekilde02}. Experimental results seems to confirm the complex behavior of oscillations in the brain activity \cite{akay09}. The main characteristic of such problems is that, although the behavior of the uncoupled units can be well understood, the collective behavior can give rise to a wealth of new phenomena hardly expected from the analysis of the isolated units. One of the archetype of the emergence of complex behavior are cellular automata \cite{wolfram86}. Cellular automata are constructed from a set of units that can be in a number of discrete states. Local rules determines the coupling between neighboring units. Certain rules, simple in their own account, can result in correlations between the units at length scales much larger than the range of the local rules and with long persistence in time. Surprisingly enough, it has been found at least one rule that results in a cellular automaton that constitutes an Universal Turing Machine \cite{wolfram02,cook04}.

Very recently Alonso \cite{alonso17} (from now on AL17) proposed a system of coupled non linear oscillators governed by a simple equation derived from Adler model \cite{adler73} that exhibits, for certain values of the control parameters, complex behaviors. The coupling is local in nature and therefore, the emergence of a rich set of different behaviors and patterns resembles in some way the same phenomena found in cellular automata. To analyze the behavior of the system in the space of the control parameters AL17 uses complexity analysis derived from the Lempel-Ziv estimation of entropy density, as has been done in the study of cellular automata \cite{ninagawa14,ninagawa14b,estevez15}. The use of Lempel-Ziv estimation in non-lineal coupled systems is not new, as it has been used before in the study of array of resonators \cite{neiman97}. Al17 study identified three distinctive regions in the space of the control parameters. One region settles into low entropy phases after a sufficient long time and is called absorbing region, a second region is termed chaotic and shows, after some time, disordered states with no spatio-temporal correlations, and a third region is identified where complex spatio-temporal patterns are observed. 

There have been arguments against using commercial compression algorithms for estimating entropy density due to the finite nature of the dictionary in compression software, it is known that for high entropy, commercial compression algorithms can significantly overestimate entropy density if the sequences analyzed are not large enough \cite{khmelev00,melchert15}. Additionally, the compression rate can deviate from a lineal relation, specially at intermediate values of entropy density. Lempel-Ziv estimation is more reliable as longer the input data, and if possible, it is always convenient to use data sequence above $10^ 4$ \cite{estevez13}. 

AL17 study of entropy focused on the analysis of the spatial-temporal behavior as a whole, not splitting the spatial entropy from the temporal entropy. If the system, as expected, resembles cellular automata type of behavior, then the entropy of temporal evolution and spatial configuration need not be simply related, and the analysis of each separately could be of interest.

In this contribution we revisit AL17 studies while expanding the analysis to other areas such as the study of the density evolution of the dynamics, and the sensitiveness to control parameters perturbation. The results will allow to explore more in detail several features already reported, as well to report new findings. The paper is organized as follows. In section \ref{sec:model} the mathematical model is formally introduced and some results from AL17 are summarized. Also the flow diagram and equilibrium conditions for the case $\gamma=0$, not studied in AL17, is presented. In section \ref{sec:methods} the methods of analysis are described. Lempel-Ziv estimation of entropy density is presented as well as the use of information distance based on the same estimates. Section \ref{sec:global} presents the results of the global analysis of the system in the phase space of the control parameters, followed in Section \ref{sec:local} by the analysis of two local transition in the phase diagram, one with constant $\omega$ and the other with constant $\gamma$. Conclusion follows.

\section{Ring of coupled non linear oscillators}\label{sec:model}

AL17 model considers $N$ oscillators arranged along a circle, where each oscillator individually complies with an Adler type equation \cite{adler73} $\dot{\theta}=\omega-\cos \theta$ for the phase, and coupled with the nearest neighbors oscillators by a non linear coupling, resulting for each oscillator in a phase evolution according to
\begin{equation}
 \frac{d\theta_i}{dt}=\omega+\gamma \cos \theta_i+(-1)^ i k \left [ \cos \theta_{i-1}+\cos \theta_{i+1}\right ].\label{eq:mastereq}
\end{equation}

\begin{figure}[!ht]
\centering
\includegraphics[scale=0.4]{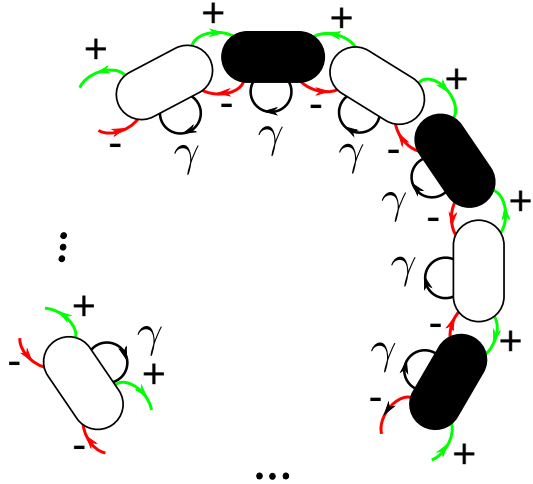}
\caption{The model is made of $N$ coupled non linear oscillators forming a ring (cyclic boundary conditions). Each oscillator is coupled with its nearest neighbors. The alternating sign of the interactions balance the coupling over the ring. An intrinsic excitation term or feedback is also considered with strength given by the control parameter $\gamma$ equal for all oscillators. A natural frequency $\omega$ equal for all the oscillators is the other control parameter. The system is described by the  equations (\ref{eq:mastereq}).
}\label{fig:osc}
\end{figure}

The state of the system is given by the set of $\theta_i$ $(i=1,2, \ldots N)$ phases, where $N$ is taken to be an even number. $\omega$ describes the natural frequency of the oscillators when no feedback and coupling is present; $\gamma$ controls the strength of the intrinsic feedback or excitation term; and the third term is the coupling with the nearest oscillators which is guaranteed to be balanced by the alternating sign, $k$ determines the strength of the coupling. Figure \ref{fig:osc} shows a diagram of the model. Circular boundary conditions are enforced by taking the index $i$ modulo $N$.

The observable of interest will not be the value of $\theta_i$ but instead the activity defined as $\sin \theta_i$.

In the case of non-coupled oscillators ($k=0$) a pure Adler behavior results.  Phase locking ($\dot{\theta}=0$) occurs for $\cos \theta=-\omega/\gamma$ which implies that equilibrium points only exist for $\gamma\geq\omega$. In the interior points of this regions there will be two equilibrium points, one of which is stable. For inner points of this region the phase gets locked at the stable equilibrium after some transient behavior that depends on the initial phase. At the boundary $\omega=\gamma$ only one equilibrium point is found. The equilibrium points are  saddle points of time. For $\omega > \gamma$, the activity shows a periodic behavior with the period depending heavily on $\omega$ and 
also on $\gamma$.

When $k\neq0$ the oscillators become coupled with their nearest neighbors.

AL17 discussed the equilibrium points for the configuration of $N$ oscillators, for $N=2$ equations (\ref{eq:mastereq}) reduce to
\begin{equation}
 \begin{array}{l}
   \frac{d\theta_1}{dt}=\omega+\gamma \cos \theta_1-2k\cos \theta_2,\\\\
   \frac{d\theta_2}{dt}=\omega+\gamma \cos \theta_2+2k\cos \theta_1,\\\label{eq:twoosc}
 \end{array}
\end{equation}
observe that in accordance with equation (\ref{eq:mastereq}) there is a factor $2$ which AL17 reabsorbs into $k$ and rescales. If we rescale  $2k\rightarrow k$  for the two coupled oscillator configuration, then the equilibrium points are given by
\begin{equation}
 \begin{array}{l}
\cos \theta_1^*=-\omega \frac{k+\gamma}{k^2+\gamma^2},\\\\  
\cos \theta_2^*=\omega \frac{k-\gamma}{k^2+\gamma^2},\\\\\label{eq:equilibrium}
 \end{array}
\end{equation}
which determines four point when 
\begin{equation}
 \omega < \frac{\gamma^2+k^2}{\gamma+k}\label{eq:eqregion}
\end{equation}
and two points at the boundary. 

\begin{figure}[!ht]
\centering
\includegraphics[scale=0.6]{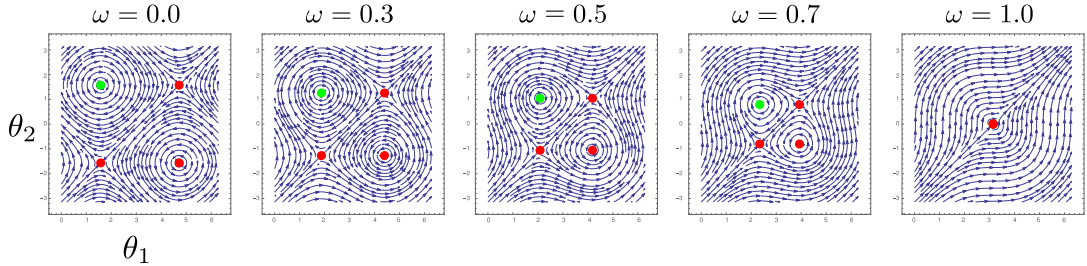}
\caption{Flow diagrams along the line $\gamma=0$ for $N=2$. The arrows indicate the direction of the flow. The equilibrium points ($\dot{\theta}_i=0$) are shown where the green one is for the point stable to phase variations, the red points are unstable. As $\omega\rightarrow 1$ the equilibrium points converge to $(\pi, 0)$ resulting in a unique equilibrium point for $\omega=1$. 
}\label{fig:flowg0}
\end{figure}

The line $\gamma=0$ was not discussed by AL17. In such case the feedback of the $i$-th equations comes through its nearest neighbors and no intrinsic excitation term or self-feedback exist. At $\omega=0$, the phase portrait is shown in the flow diagram of figure \ref{fig:flowg0}(right). Four equilibrium points exist. Of this four points two are unstable with respect to a variation of $\theta_1$ and $\theta_2$. The other two equilibrium points are quite interesting. The regions around such points determines non intersecting orbits circling around the equilibrium points. For any pair of initial  $(\theta_1, \theta_2)$ near enough to one of these two points, the system gets trapped indefinitely in such orbits which do not get closer nor farther to the nearest equilibrium point: they are neither repulsive nor attractive.  When $\omega\rightarrow 1$ the four equilibrium points start converging to a single unstable equilibrium at $(\pi,0)$ which is attained at $\omega=1$ (figure \ref{fig:flowg0})(left).

For more than two oscillators, while keeping $N$ even, the region with equilibrium points is still determined by the inequality (\ref{eq:eqregion}) if the described rescaling $2k\rightarrow k$  is performed.

The system can be further rescaled in terms of $k$ (this merely changes the time units) which allows to consider, from now on, $k=1$ in equation (\ref{eq:mastereq})  without loss of generality. 

\section{Data processing and complexity analysis procedure}\label{sec:methods}

In order to perform the analysis of the behavior of the spatio-temporal evolution of the coupled oscillator in terms of complexity analysis, first a discretization of the output signal (the so called phases activity: $\sin \theta_i$) has to be done. There are two usual ways to binarize data for such purposes. In one case the mean value of the output spatial data is determined and all values above or equal to the mean are taken as one value, say $1$, while the others  take the value $0$. This procedure does not guarantee the same number of $1$s or $0$s, reduction of complexity can be result of the production of one digit at the expenses of the other or, as a shuffling of the digits in the spatial data resulting in the formation of patterns. In the second procedure for binarization, the median value is taken as the threshold. The use of the median fixes, in any typical run, the occurrence of both digits at half the total number of characters in the output string ($N/2$). In this case, any variation of complexity is result solely of the shuffling of digits in the binary string. In this work the first procedure will be used to follow the analysis made in AL17. The use of the mean allows to consider the fraction of the number of $1$s in the binarized string as an observable variable which will be called the density and denoted by $\rho$. The variation of the density can be analyzed in terms of Landauer principle which, in this case, means that any change in density can be associated with a dissipation of energy \cite{landauer61}.  

After binarization the spatial configuration of the oscillators will be described by a binary string $s$ of length $N$. Simulations were performed over $N=5 \times 10^3$ coupled oscillators for the density calculations and $N=10^4$ for all phase diagrams involving calculations of entropic magnitudes. For specific values of $(\omega, \gamma)$, systems of up to $N=2\times 10^5$ oscillators were used. Averaging over $100$ runs of random initial conditions were performed. The system of differential equations given by (\ref{eq:mastereq}) was solved numerically using a Runge-Kutta method of order $4$ \footnote{Runge-Kutta method was used as implemented in the GNU Scientific Library (GSL) and in the Mathematica system (Wolfram Inc). Calculations made in both systems were compared for several values to discover any discrepancy in the stability of the solutions, while the final calculations were made, for speed reasons, with GSL.}.

\subsection{Lempel-Ziv estimate of entropy density and information distance}\label{sec:lz76}

There are different flavors of Lempel-Ziv complexity and the reader must take care in the one thats been used.  In all cases it is based on a factorization of a string according to certain rules. The analysis in this contribution will be based on the factorization described in \cite{lz76}.

The ''drop'' operator $\pi$ acts over a string by removing a character at the end of the string. $s(i,j)$ denotes a substring starting at position $i$ and ending at position $j$ (length $j-i+1$) then 
\begin{equation} 
s(i,j) \pi=s(i,j-1). \nonumber
\end{equation}
Any integer power of the drop operator can be interpreted recursively,
\begin{equation}
s(i,j) \pi^k=(s(i,j)\pi)\pi^{k-1}=s(i,j-1)\pi^{k-1}=s(i,j-2)\pi^{k-2}=\ldots=s(i,j-k). \nonumber
\end{equation}

Consider a factorization $F(s)$ in $m$ factors of the string $s$
\begin{equation}
F(s)=s(1,l_1)s(l_1+1,l_{2})\dots s(l_{m-1}+1,N), \nonumber
\end{equation}
such that each factor $s(l_{k-1}+1,l_k)$ complies with
\begin{enumerate}
 \item $s(l_{k-1}+1,l_k)\pi\subset s(1,l_k)\pi^2$
 \item $s(l_{k-1}+1,l_k)\not\subset s(1,l_k)\pi$ except, eventually, for the last factor $s(l_{m-1}+1,N)$.
\end{enumerate}
This factorization is called the Lempel-Ziv factorization of the string and it is unique \cite{lz76}.

For example the sequence $u=01001011101$ factorizes as $F(s)=0.1.00.101.11.01$, where each factor is delimited by a dot.

The Lempel-Ziv complexity $C_{LZ}(s)$ of the string $s$, is defined as the number of factors  $(=|F(s)|)$. In the example above, $C_{LZ}(s)$=6.

The Shannon entropy density is defined by 
\begin{equation}\label{shannonh}
 h(s)= \lim_{N\rightarrow\infty}\frac{H[s(1,N)]}{N},
\end{equation}
where $H[s(1,N)]$ is the Shannon block entropy of the string $s(1,N)\in s$  \cite{cover06}. Ziv theorem \cite{ziv78} states that $h(s)$ can be estimated by
\begin{equation}
  h=c_{LZ}=\limsup_{N\rightarrow\infty} \frac{C_{LZ}(s)}{N/\log{N}}.\label{eq:hlz}
\end{equation}
Lempel-Ziv complexity for long enough strings can be used as an estimate of the entropy density \cite{lesne09,estevez13}.

Available compression algorithms such as gzip, bzip or other commercially available softwares have been used as estimates of the Lempel-Ziv complexity of a string \cite{emmert10,alonso17}. In general care must be taken with the use of such software for entropy estimation and moreover, there is no need for such procedure. There are several drawbacks of using compression ratios as an estimate of $C_{LZ}$. For those compressors based on the Lempel-Ziv algorithm, the factorization used is not the one described above, but instead one based on another procedure described in \cite{lz77}. Both procedures are not equivalent. More importantly is the fact that in compressors software a finite size dictionary of factors is used, this means that if, while factoring a sequence, the number of factors exceeds the fixed size of the dictionary no further factors are added. Lempel-Ziv complexity of the string, upon a certain value, starts not to be linearly proportional to the compression ratio and the latter turns to be not a good estimate \cite{khmelev00,melchert15}. On the other hand, if short sequence are used for factorization the estimate given by (\ref{eq:hlz}) can be seriously affected by systematic errors \cite{schurmann99,lesne09}. The convergence speed of equation (\ref{eq:hlz}) is extremely slow \cite{estevez13}. Fortunately there is no need to reach for compression softwares to estimate $C_{LZ}$, there are exact Lempel-Ziv factorization algorithms with an efficient economy in time and memory that can be as fast as $O(N)$\cite{crochemore86,chen08}.

Lempel-Ziv complexity can be used to define an information distance between two sequences. The idea of information distance based on Kolmogorov randomness was introduced by  Li et al. \cite{li04}. This definition was adapted by using compression algorithms as estimates of Kolmogorov complexity by Li et al itself and by Emmert-Streib \cite{emmert10}. In \cite{estevez15} Lempel-Ziv complexity was used to derive an information distance. The expression that will be considered as the information distance between two string $s$ and $p$ of equal length is given by,
\begin{equation}
 d_{LZ}(s,p)=\frac{C_{LZ}(sp)-min\{C_{LZ}(s), C_{LZ}(p)\}}{max\{C_{LZ}(s), C_{LZ}(p)\}}.\label{eq:dlz}
\end{equation}
where $sp$ means concatenation. The rationale behind expression (\ref{eq:dlz}) is discussed in \cite{estevez15}, the difference between (\ref{eq:dlz}) and the expression used there, is that here the Lempel-Ziv complexity $C_{LZ}$ is used instead of the Lempel-Ziv estimate of the entropy density to account for the finiteness of the involved strings.

The distance given by equation (\ref{eq:dlz}) does not quantify the number of bits where the two strings differ, instead is a measure of how much factors one string has in common with the other and therefore, characterize, from the Lempel-Ziv factorization procedure, how innovative one string is seen once the factor of the other string are known. In this sense it is more an information distance than a quantification associated to the damage field, that is the Hamming distance between two strings.

Calculations of both Lempel-Ziv complexity and information distance were made using an in-house software, already used in the study of cellular automata \cite{estevez15} which runs in $O(N \log N)$ time.

\section{Global analysis of the control parameter space: phase diagram analysis}\label{sec:global}

Figure \ref{fig:clz}(spatial) reproduces roughly, by our own calculations, the phase diagram of AL17. The $c_{LZ}$ estimates of entropy density was calculated from the final oscillator configuration. As it will be seen, all phase diagrams drawn from different measures have qualitatively the same form. A region in black in the phase diagram is found for those combinations of $(\omega, \gamma)$ values that result in zero entropy density and it covers approximately half of the diagram. There is a wedge region in green, with intermediate value of entropy density. A needle region, also in green, is embedded within a mostly red region of high entropy density, the latter with values near the maximum value of $1$. The low entropy region was called absorbing in AL17 while the label of chaotic was reserved for the highest entropy density region (Figure \ref{fig:clz} diagram). In the chaotic region, the calculations shows some disturbance of lower entropy density, not reported in AL17 phase diagram. This region does not define a connected area and we called it ``turbulent region''. 

Entropy density was also calculated for the temporal evolution. In this case the binarized string represents the activity of a single oscillator over time. Temporal entropy density as estimated by $c_{LZ}^ t$ was calculated from such string. Averaging in this case is made over the $5000$ oscillators. Figure \ref{fig:clz}(temporal) shows the obtained phase diagram. Several difference with the spatial diagram can be found. First, the wedge intermediate spatial entropy density region, as well as the needle region now exhibit almost zero entropy density. The chaotic region is not as homogeneous as seen in the spatial phase diagram. The entropy density is higher near the border of the chaotic region and decreases for increasing value of $\omega$. The turbulent region within the chaotic region is now more clearly seen as green values below $0.5$ but above zero. 

Calculations of $c_{LZ}$ over the final configurations have the drawback that it is assumed implicitly that the system settles with time to some ''final'' configuration. As discussed for cellular automata \cite{estevez15} this has not to be the case and the system can have cycles of configurations with more or less long periods. Temporal analysis avoids such assumption. On the other hand, there is no reason to assume that, for a given set of control parameters the value of spatial configuration complexity has to be similar to the temporal calculation. Our simulations shows that this is not clearly the case. Spatial entropy density measures the possible lack of correlation and the lack of pattern in the coupled oscillators at a particular value of time, while temporal entropy density measures the same for a single oscillator along a time-line. Even if they need not to be similar, one should expect some relation between spatial and temporal entropy which accounts for how much disorder in the temporal correlations can be accommodated by a given disorder in the spatial correlations.

\begin{figure}[!ht]
\centering
\includegraphics[scale=1.5]{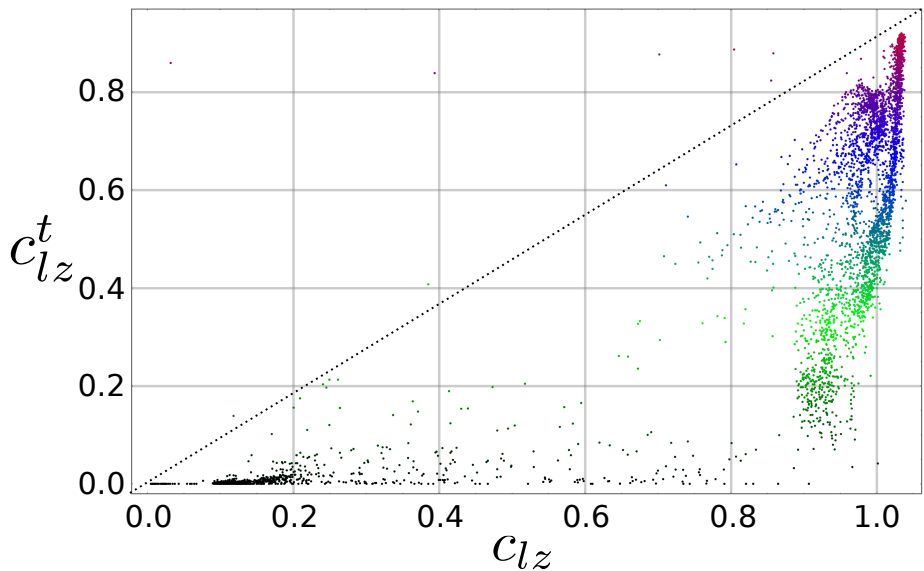}
\caption{Entropy estimates $c_{LZ}^ t$ from the temporal evolution  as a function of the entropy estimate  $c_{LZ}$ from the final spatial configuration. Data points were drawn from the phase diagrams of figure \ref{fig:clz} and the same conditions in the simulations apply.
}\label{fig:spat_vs_temp}
\end{figure}

Figure \ref{fig:spat_vs_temp} is a plot of the temporal entropy density as a function of the spatial entropy density. It is clear that low values of spatial entropy density can only result in low values of temporal entropy density. Yet, as $c_{LZ}$ increase, the system can accommodate values of $c_{LZ}^ t$ below the line $c_{LZ}^ t=c_{LZ}$. Consider the wedge and needle regions, although they exhibit intermediate values of $c_{LZ}$, their temporal complexity density is zero showing that either a final constant in time state of the oscillator is reached or a periodic one, yet spatial configurations are far from homogeneity or periodic correlations. Within the chaotic region, temporal homogeneity or periodicity is absent. There are points at the boundaries of the chaotic region, where the temporal entropy density is almost one. For such points it could be hard to talk of a ''final'' configuration. For points with lower $c_{LZ}^t$ but still clearly above zero, some patterns could be potentially discovered in the temporal evolution but randomness is still much a property of the time evolution, again, care must be taken to define any ''final'' configuration.

\begin{figure}[!ht]
\centering
\includegraphics[scale=0.7]{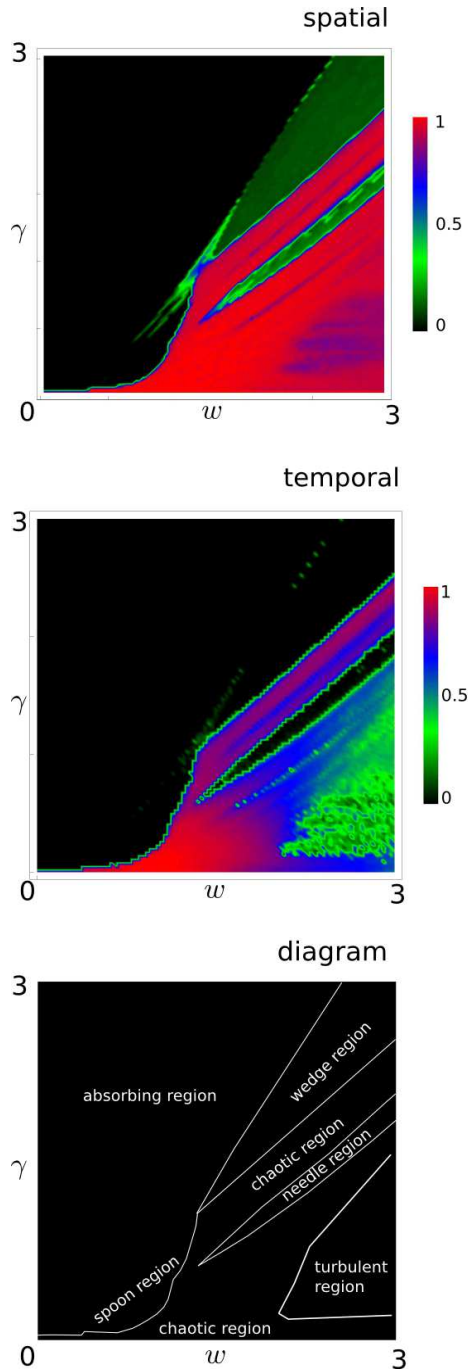}
\caption{The entropy density estimates from the Lempel-Ziv complexity over the control parameters $(\omega, \gamma)$ space. In all cases the number of oscillators is $N=10^4$ and $10^4$ time steps are taken. (spatial) Corresponds to the $c_{LZ}$ value from the final configuration of the oscillators after $10^4$ steps. Each point is the average of $100$ runs. (temporal) Corresponds to the $c_{LZ}^t$ value from the temporal evolution of each oscillators. Each point is the average value for all $10^4$ oscillators in the ring.  The step in both control parameters is $0.03$. (diagram) A schematics of the different regions identified in the two maps above.
}\label{fig:clz}
\end{figure}

\begin{figure}[!ht]
\centering
\includegraphics[scale=0.5]{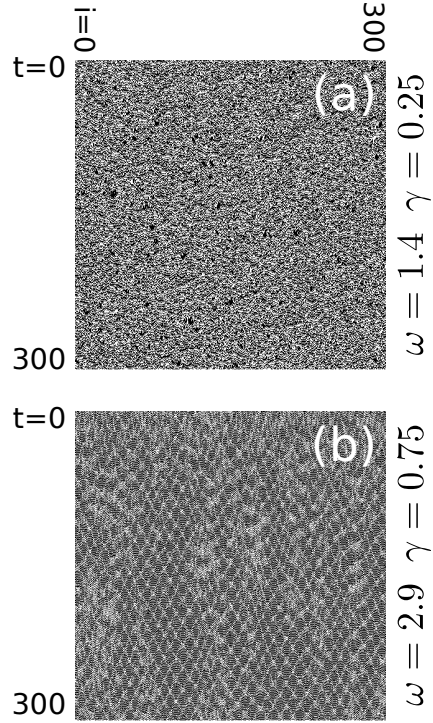}
\caption{Spatio-temporal pattern for a point in the (a) chaotic region and, (b) turbulent region. The vertical axis is the temporal evolution starting at the upper line and evolving downwards. A disturbed periodic pattern can be seen in the diagram of the turbulent region point accounting for the decrease of temporal entropy density seen in the middle plot of figure \ref{fig:clz} 
}\label{fig:turbulent}
\end{figure}

By splitting the entropy analysis into the spatial configurations and the temporal evolutions it can be seen that the chaotic region is not as homogeneous as reported by AL17 (See figure 2A and 3A in \cite{alonso17}). As mentioned earlier, there  is a ''turbulent'' region identified in green in figure \ref{fig:clz}(temporal) within the chaotic region, with clearly lower entropy rate,  that was not reported in AL17. In this region the spatial phase diagram (Figure \ref{fig:clz} spatial) exhibits a slightly smaller entropy rate. What both phase diagrams are signaling is that the entropy behavior in the chaotic region can have different nature. While in the whole of the chaotic region the spatial entropy rate of the final configuration has the maximum value of $1$ or is very near to it, the temporal evolution can have different behaviors. Outside the turbulent region, temporal patterns in the individual oscillators evolution can not be found. There is no final spatial configuration in terms of a fixed, periodic or other type of ordering in the spatial configurations time evolution. Inside the turbulent region, on the other hand, temporal entropy rate is lowered significantly ($<0.5$) which means that the system evolution acquires patterned behavior to some extent, even if it is not perfect. The highly disordered spatial configurations, evolves obeying some law, be it fixed, periodic, quasi-periodic or of some other patterned nature. Figure \ref{fig:turbulent}b shows the spatio-temporal diagram of a point within the turbulent region. After an initial transient the system exhibits a disturbed periodic pattern which contrast with figure \ref{fig:turbulent}a, taken from a point in the chaotic region,  where no such patterned diagram can be observed in the whole time evolution. Further investigation of the turbulent region is still underway and will be reported elsewhere. 

\begin{figure}[!ht]
\centering
\includegraphics[scale=1]{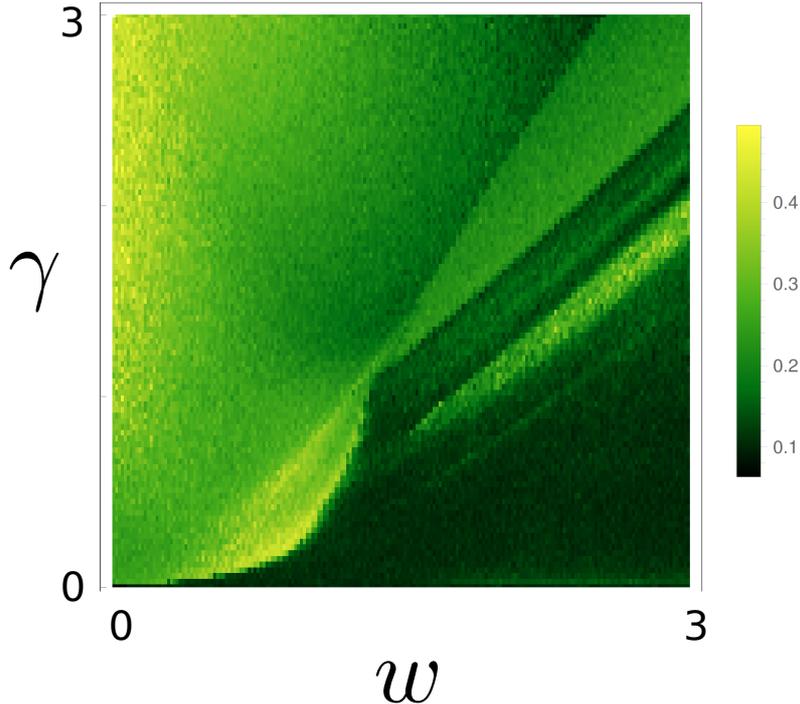}
\caption{The maximum of the absolute value of the variation of density values from the initial $1/2$ value: $|\rho-1/2|_{max}$ for all time steps over the control parameters. The conditions in the plot are the same as in figure \ref{fig:clz}.
}\label{fig:density}
\end{figure}

Figure \ref{fig:density} shows the phase diagram for the maximum density variation. The initial value of density for the binarized string of oscillators activity is in all cases $1/2$, the largest absolute value of deviation of the density from this value is used in figure \ref{fig:density}.  The absorbing region is in general dissipative with the largest value of density variation for the lower values of $\omega\sim 0$. The density variation decrease as the chaotic region is approached. In those regions with large dissipation and low entropy density as can be found in the absorbing region, the reduction of randomness comes as a result mainly of the production of one digit at the expenses of the other. As the production of one symbol at the expense of the other involves irreversible erasure, Landauer principle implies that time evolution, at some point, involves large variation of the energy of the system, thus the name of dissipative. Values of density variation near the border of the absorbing region to the wedge can exhibit low values, even near zero.  For such values of the control parameters, reduction of entropy density comes as a result of the ordering of the digits in the binary sequence and not by irreversible erasure actions, reduction of entropy in such case could be reversible. 

The wedge region has intermediate value of density variation and its clearly distinguished in the phase diagram between the absorbing and the chaotic region. Reduction of entropy density is a mixture of digit erasure and digit reordering in the spatial configuration.  Dissipation is again high for the needle region and therefore heavy erasure actions have taken place at some point in the time evolution. In the chaotic region dissipation attains it lowest value which together with the high entropy density could be an indication of little changes in the time evolution of the binary string. In the density variation analysis a new region emerges in what in AL17 was called the spoon feature of the phase diagram, it has a straight boundary from above and a concave spoon like shape from below. the entropy density, both spatial and temporal, is zero within this region and the density variation has a jump, specially from below, from its neighbors regions. 

Calculations were also performed for the analysis of the stability of the system to perturbation in the initial conditions and small variations in the control parameters. 

\begin{figure}[!ht]
\centering
\includegraphics[scale=1]{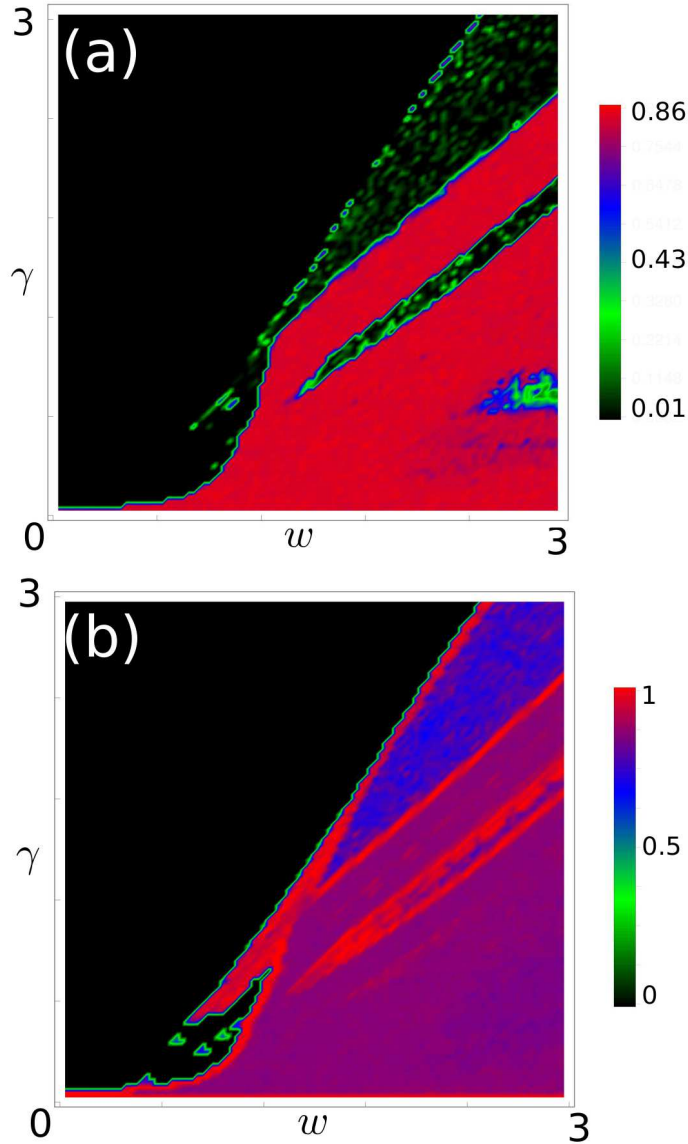}
\caption{The information distance between two runs over the control parameters. In (a) the two runs corresponds to the same initial conditions except at one oscillator. The plot then measures the divergence of behavior after $10^4$ steps starting from initial information distance of almost zero. In (b) the $d_{LZ}$ value corresponds to the maximum information distance between two runs from nearest neighboring control parameters $(\omega\pm0.03, \gamma\pm0.03)$. The conditions in the plot are the same as in figure \ref{fig:clz}.
}\label{fig:perturb}
\end{figure}

Figure \ref{fig:perturb}a shows the information distance $d_{LZ}$ between configurations at a time step of $10^4$ that started from identical initial conditions except for a single oscillator ($N$ was taken as $5\times 10^ 3$). Information distance measures the departure of one system behavior with respect to the other. As expected, in the absorbing region, the system is insensible to small variations of the initial conditions. The same is true in the wedge and needle region as well as the spoon region. A much heavier dependence on initial conditions is found in the chaotic region.

The analysis performed using LZ-distance differs from that using standard deviation measure: two different initial conditions can result in similar entropy density and yet their information distance could be large. AL17 study measures, using standard deviation of entropy over a set of initial conditions,  how sensitive is the entropy measure to variations of initial conditions. Both analysis are complementary. It is therefore no surprise that the phase diagram that emerges from both measures area different. During our numerical experiments it was found that using the same values $(\omega, \gamma)$ over two initial configurations could lead to patterns with similar entropy density and yet far apart from a information distance point of view.  It should be emphasized what does the LZ-distance measures. LZ-distance quantifies how much of the pattern (factor) production of one sequence can be found in the other sequence. A large information distance means that the patterns (factors) found in one sequence are not very informative on the patterns (factors) found in the second sequence. In that sense, if two sequence with similar entropy density have an information distance near one, then one can say that the patterns found in one sequence are not related by the patterns found in the other one, in spite of both having the same amount of irreducible randomness. 

Figure \ref{fig:perturb}b is the diagram drawn from the perturbation of the control parameters. In this case the same initial condition was taken for all values of the control parameters and the final configurations were compared to the nearest neighboring points $(\omega, \gamma)$. The maximum $d_{LZ}$ among all near neighbors comparison was taken. A picture with substantially difference with the   previous one emerges. The absorbing region is still quite insensible, which now is interpreted as being stable to the variation of the control parameters. But now the wedge region is much more sensitive (unstable) to the control parameters. The chaotic region is very unstable to the value of the control parameters. The needle region, which was insensitive to the initial conditions, seems to be sensitive to small variations of the control parameters, a result that is associated to the narrow geometry of the region.  All borders between different regions are highly sensitive to perturbations of the control parameters.

\section{Local analysis of the control parameters space}\label{sec:local}

\subsection{Transition from the absorbing region into the chaotic region along a line of constant $\omega$}\label{sec:vert}
\begin{figure}[!ht]
\centering
\includegraphics[scale=0.55]{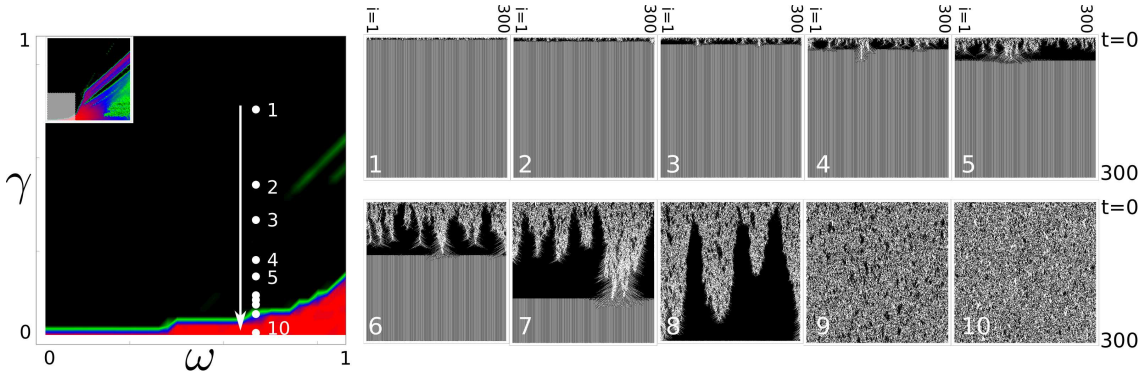}
\caption{\textbf{Left}: The transition from the absorbing state to the chaotic region along a line of constant $\omega=0.70$, starting at $\gamma=0.75$ to $\gamma=0$. \textbf{Right}: The spatio-temporal evolution of the oscillators binary strings at the different points shown at the left.
}\label{fig:vert}
\end{figure}

Consider the behavior of the system as it moves from the absorbing region at $\omega=0.70$, $\gamma=0.75$ to the chaotic region at  $\gamma=0.0$ along the line of constant $\omega$ shown in figure \ref{fig:vert}(left). In all cases the initial configuration is random. In the absorbing region for the largest values of $\gamma$, the system rapidly settles in a configuration that does not vary with time (phase locking) and at the final step, a low entropy density configuration can be identified. The locked configurations is made of alternating $0$s and $1$s. As $\omega$ decreases the time needed for the system to reach phase locking increases, as a result, two regions are identified in the spatio-temporal diagrams ( See for example figure \ref{fig:vert}(right)-6).  The first region is the immediate evolution of the initial random state and no phase locking happens, which can be seen from the evolving patterns. Within this initial region, patches of heavily disordered regions remanent of the initial random state shrink with time and in some places resembles dendritic patterns. At the triggering time value the system suddenly changes its spatial configuration to a low entropy phase (of alternating $0$'s and $1$'s) and locks its phases signaling the start of the second region. The systems changes to the behavior found in the deep absorbing region. The smaller the $\gamma$ value, the longer it takes to reach the triggering event which can be seen in the succession of spatio-temporal diagram of Figure \ref{fig:vert}(right).  The analysis of the spatial entropy density is not able to follow this behavior as it only looks into the final configuration that in all this cases will have similar values of $c_{LZ}$, the same limitation pertains the analysis of the spatio-temporal evolution as a whole, as done in AL17. Time evolution is much richer that this two analysis show. 

\begin{figure}[!ht]
\centering
\includegraphics[scale=1.2]{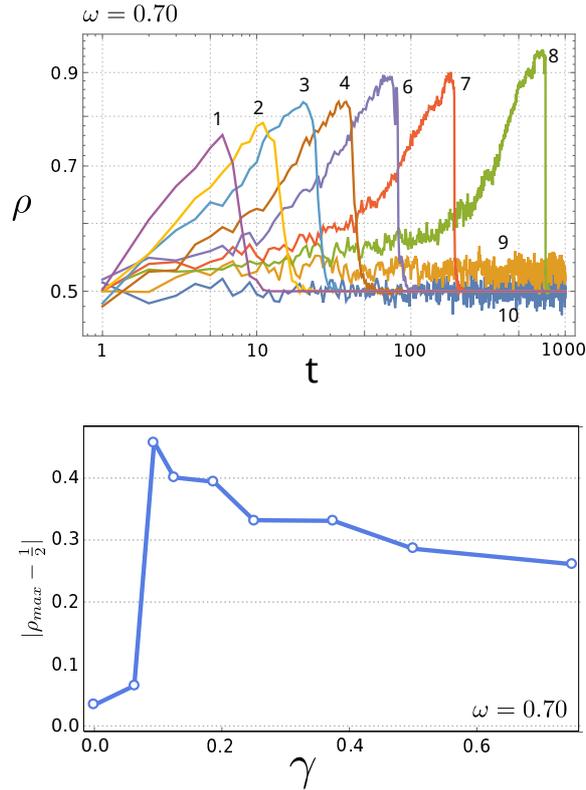}
\caption{\textbf{Up}: The evolution of density with time at the different points described in figure \ref{fig:vert} along the line with $\omega=0.70$. In all cases the configuration starts at a value of $\rho=0.5$. \textbf{Low}: Maximum absolute value of he density deviation from $0.5$ as a function of $\gamma$ for $\omega=0.70$. 
}\label{fig:vertdens}
\end{figure}

Figure \ref{fig:vertdens}(up) shows that within the first initial region, strong production of $0$s digits is attained. This, in turn, shows up in the spatio-temporal diagrams of figure \ref{fig:vert}right as the solid black regions.  A steady departure of the density from the value of $0.5$ as time increases happens with constant slope followed by an  abrupt fall to $0.5$ at a specific triggering value of time which depends on $\gamma$. The reader should notice that within the first region, as time evolves, the erasure of $1$s causes a decrease of entropy density. The departure of the density from the initial value of $0.5$ has a maximum for $\gamma=0.093$ and from there starts decreasing with increasing $\gamma$ value (Figure \ref{fig:vertdens} low). The random regions that split into patches as time evolves hardly survive at the time when the triggering occurs. It seems as if triggering to phase locking needs the erasure of the random regions before it can happen. Such erasure happens faster for larger $\gamma$ values. 

For $\gamma < 0.06$ (Figure \ref{fig:vert}(right)-9,10) there is no triggering value. No steadily production of $0$s ever happens and the density has small oscillations around $0.5$ with a reshuffling of the oscillators values that does not create any recognizable pattern. The system has a random behavior both spatially and temporal. Phase locking is not attained. As no spatial correlation can be identified, there are no evidence of local collective behavior of the oscillators except (Figure \ref{fig:vert}(right)-9), for some short-time small islands of $0$s within the random region. The ephemeral black islands seems uncorrelated between them and resembles more like a fluctuation phenomena. The $0$s islands have further shrink size for $\gamma=0$ and are hardly recognizable in figure \ref{fig:vert}(right)-10.

The above behavior can be explained in terms of the equilibrium points. At larger values of $\gamma$ the system quickly settles into the stable equilibrium point and phase locking occurs. This equilibrium point is the alternating configuration $101010\ldots$ of the configuration string. As $\gamma$ decreases the system spends more time to settle into the stable equilibrium point. The wandering of the system decays to the equilibrium point by circling it in longer and longer periods. Finally at $\gamma=0$, the system could gets trapped into an ever lasting circling around the equilibrium point as was shown for $N=2$ in the fourth flow diagram of figure \ref{fig:flowg0}. Phase locking is never attained.

\subsection{Transition from the absorbing region into the chaotic region along a line of constant  $\gamma$}\label{sec:horz}

\begin{figure}[!ht]
\centering
\includegraphics[scale=0.61]{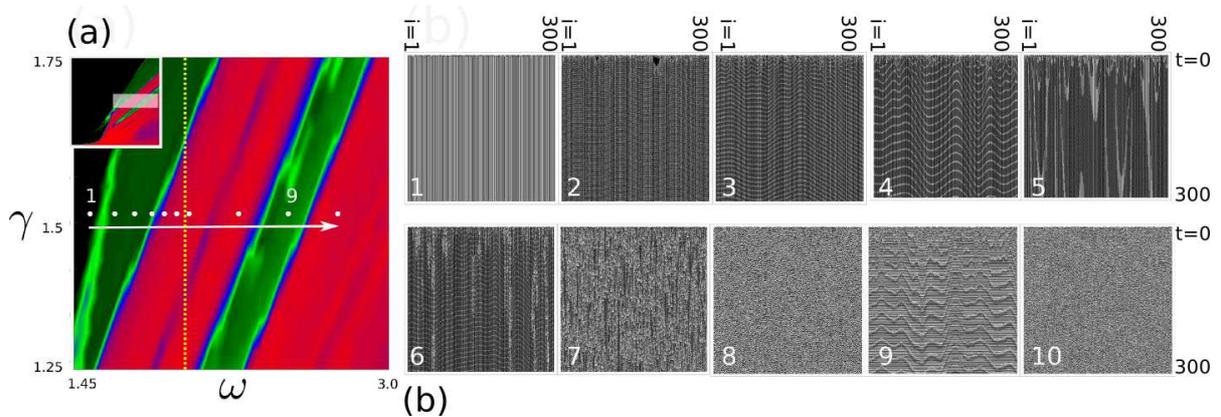}
\caption{\textbf{(a)}: The transition through several regions along a line of constant $\gamma=1.5$, starting at $\omega=1.5$ to $\omega=2.75$. The vertical dashed line marks the border between the region with equilibrium points at the left, and where no equilibrium points are possible to the right. \textbf{(b)}: The spatial-temporal evolution of the oscillators binary strings at the different points is shown at the left.
}\label{fig:horiz}
\end{figure}

\begin{figure}[!ht]
\centering
\includegraphics[scale=0.7]{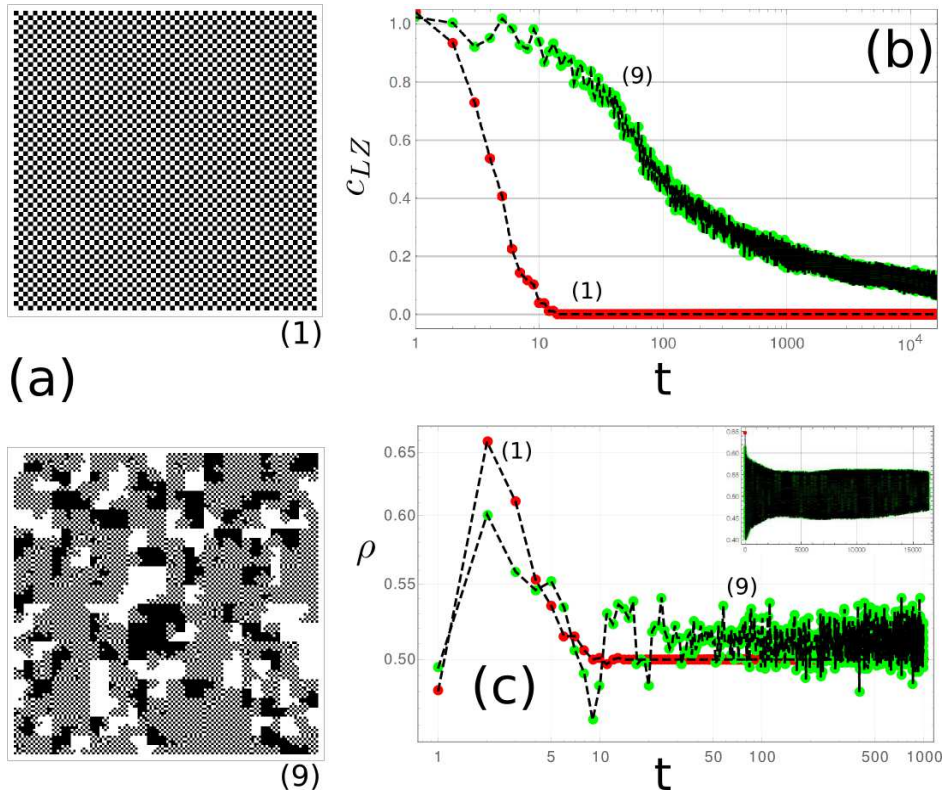}
\caption{\textbf{a}: Hilbert plot of the spatio configuration at $(\omega,\gamma)=(1.5,1.5)$ (up) and $(\omega,\gamma)=(2.5,1.5)$ (down), corresponding to point $1$ and $9$ of figure \ref{fig:vert}, respectively; \textbf{b}: The temporal evolution of the entropy density estimate $c_{LZ}$ at points $1$ and $9$;  \textbf{c}: The temporal evolution of density at points $1$ and $9$.
}\label{fig:horizclz}
\end{figure}

There are three regions in the transition pictured in figure \ref{fig:horiz}a. The first point corresponds to the absorbing region, low entropy, phase locked configurations can be seen. The configuration is of alternating $1$s and $0$s as can be seen in the check-board pattern of the Hilbert plot \footnote{Hilbert plot is a local preserving mapping of a one dimensional array into a two dimensional arrangement as described in \cite{estevez15a}. It allows a compact visualization of long runs of data while preserving neighboring features of the original sequence.} of figure \ref{fig:horizclz}a (up). In the wedge region , point $2$, $3$ and $4$ of figure \ref{fig:horiz}a, less trivial patterns can be found where each oscillator gets trapped in a periodic orbits for sufficiently long times. From $2$ to $4$, the wave like nature of the patterns show spatial-temporal correlation between the oscillators, as the amplitude of the waves gets larger. Point $5$ no longer exhibits a wavy nature, but there are clearly correlations in the spatial configuration and time evolution. Point $5$ is at the border between the wedge and the chaotic regions. At this point,  some oscillators shows constant values for a a period of time. This behavior can happen at few neighboring oscillators which propagates with time to neighboring oscillators at one side and eventually collides with other constant value regions, and then starts shrinking to disappearance. In one case, constant value happens at one oscillator after some time and from there spreads to neighboring oscillators at both sides. Point $6$ recovers the wavy behavior, but in this case, the wave pattern is disturbed by propagating structures along the time line. Starting  from point $7$, subsequent points are in the region where no equilibrium point exist. Points $7$ and $8$ are inside the chaotic region and no wave pattern can be recognized but clearly there are spatio-temporal correlations which seems more persistent in time for the first point. Point $9$, at the center of the needle region recovers the wavy nature found on the wedge and seems to remain there for large enough time steps, no phase locking happens but a periodic orbit of each oscillators. In this case, the final configuration is made of patches of constant value as seen in the Hilbert plot of figure \ref{fig:horizclz}a (down). Point $10$ also exhibits spatio-temporal correlations as the map at the right of figure \ref{fig:horiz}b is far from a salt and pepper contrast. 

In figure \ref{fig:horizclz}b, the time evolution of the entropy density is shown in both, point $1$ at the absorbing region ($\omega=1.5$) and point $9$ at the needle region ($\omega=2.5$). At point $1$, $c_{LZ}$ quickly decays to zero value and remains constant in correspondence with the spatio-temporal map showing phase locking. The needle region on the other hand, has a slowly decaying entropy density with strong oscillations that seems eventually to reach values near zero. Both points exhibit a rise of the density for the first time steps followed by a decay that asymptotically goes to $0.5$ (Figure \ref{fig:horizclz}c). In the case of point $1$ this is certainly the case, yet for point $9$ in the needle region, strong oscillations can again be seen and the asymptotic value seems only certain in average.

\section{Discussion and conclusions}\label{sec:concl}

The main point of this contribution is to show that there is a rich set of  behaviors for the model presented which can be effectively explored by the use of different measures. This extends the analysis made in \cite{alonso17} further confirming some of the findings made there while showing additional results. The analogy with cellular automata behavior has been brought up. The model presented has continuous, real valued, control parameters and in consequence, an infinite non-numerable number of ''rules'' happens. This is the case even for any subset of the explored region, no matter how small it is chosen. This is in contrast with the numerable nature of cellular automata rules. Cellular automata have been considered computational system capable to process and store information, the system studied could be seen under such perspective.

In cellular automata analysis, the possibility of complex behaviors is usually associated with intermediate values of entropy density. Rules with complex behavior have been studied as the more interesting candidates for sophisticated computational capabilities.  Two main regions with intermediate entropy density have been identified and were called the wedge and needle region. Other regions where complex behavior could be expected are at the boundaries between those two regions and the chaotic region, where $c_{LZ}$ shows values above $0.5$ but below $1$. The analysis shows that care must be taken to reduce entropy analysis to the spatial configuration, as temporal evolutions can show interesting features not captured by the former. The relation between spatial and temporal entropy density is by no means trivial. 

Sensitivity to initial conditions is high, as expected, in the chaotic region which implies that, in this region, the system is to unstable to any perturbation as to make it suitable for any ''useful'' processing capability. In the absorbing region, the insensitivity to initial conditions also hinders the capability to process information as a wide range of initial conditions leads to the same outcome after a short period of time. The wedge and needle region are less sensitive to the initial conditions that the chaotic region and slightly more sensitive that the absorbing states. Even more interesting is the boundary regions were sensitiveness to initial condition have intermediate values, which points to the possibility to ''harvest'' the information processing capabilities of the system at these points. The system stability to small changes in the control parameters becomes relevant, an issue that is not found in cellular automata analysis. The regions with the highest instability are the boundary regions, this is lower for the wedge region but still above $0.5$. In the needle region its narrow nature makes the more stable region to be an even narrower similar shaped area within the region that our numerically studies did not explore fully. The chaotic region has, as expected, a high instability to changes in the control parameters, but interesting enough lower than the described boundaries.

Density analysis has been useful to characterize the system in terms of its dissipative nature through Landauer principle. With respect to the density, the absorbing state is the most dissipative being the states where the systems performs the largest ordering of the initial configuration. The interesting feature here is that even if the final configuration has the same density as the initial random conditions, the systems starts by a strong production of one symbol at the expense of the other (erasing action) which attains a peak depending in the value of ($\omega$, $\gamma$) and then falls again to the $1/2$ initial value. Ordering is then found to occur, not as a simple reshuffling of symbols in the initial string, but instead through dissipative processes: the ordering capability of the system, in the absorbing state, is not by information processing but instead by information destruction and pattern creation from a highly erased configuration. In thermodynamic terms, it seems more like a first order transition with latent heat, than a higher order transition.  

The discretization procedure can be considered as a coarse-graining process where information is necessarily lost compared to the original data. As a result, discretization can result, in certain cases, in missing  meaningful aspects of the original signal, this has been analyzed for example, in the case of electroencephalogram data \cite{bali15}. With that in mind, AL17 analysis as well as the one performed in this contribution, in spite of the wealth of interesting phenomena described, could be missing additional features relevant to the complexity analysis of the data, which are lost during binarization. Further studies in this direction are therefore needed and in this context, permutation Lempel-Ziv analysis could be an interesting path to follow\cite{bali15}.

It is known that entropy analysis alone is not sufficient to asses complex behavior. In some cases, surrogate data analysis by randomizing the original data has been used to complement the entropy rate analysis  \cite{nagarayama02}. Although such analysis is still to be made for the Alonso model and will be the subject of further research. it must be pointed out that the study performed by AL17 and in this work, does not rest entirely in estimates of entropy rate. The observation of the spatio-temporal diagrams shed light in a qualitative sense to the appearance of long range and long time correlations among the oscillators.

\section{Acknowledgments}

This work was partially financed by UH institutional project LEMPELZIV2017.


\end{document}